# Emotion Specification from Musical Stimuli: An EEG Study with AFA and DFA

Sourya Sengupta, Sayan Biswas, Sayan Nag
Department of Electrical Engineering
Jadavpur University
Kolkata, India

Shankha Sanyal, Archi Banerjee, Ranjan Sengupta, Dipak Ghosh
Sir C.V. Raman Centre for Physics and Music
Jadavpur University
Kolkata, India

***Abstract*—The present study reports interesting findings in regard to emotional arousal based activities while listening to two Hindustani classical ragas of contrast emotion. EEG data was taken on 5 naïve listeners while they listened to two ragas – Bahar and Mia ki Malhar which are conventionally known to portray contrast emotions. The EEG data were analyzed with the help of two robust non-linear tools viz. Adaptive Fractal Analysis (AFA) and Detrended Fluctuation Analysis (DFA). A comparative study of the Hurst Exponents obtained from the two methods have been shown which shows that DFA provides more rigorous results compared to AFA when it comes to the scaling analysis of bio-signal data. The results and implications have been discussed in detail.**

## I. INTRODUCTION

Can music be defined? Most of us may agree that it is composed of a complex time series comprising of the three fundamental bases of physics - frequency, amplitude and timbre, but in real life music is something much more than that. The physics of music is as interesting because of its aesthetic beauty as well as its ability to convey a variety of moods through its rendition. Music signals are said to possess a chaotic but self-similar structure in different scales. At first sight music shows a complex behavior: at every instant components (in micro and macro scale: pitch, timbre, accent, duration, phrase, melody etc.) are close linked to each other [9]. All these properties (above stated in a heuristic characterization) are peculiar of systems with chaotic, self-organized, and generally, non-linear behavior. Therefore, the analysis of music using linear and deterministic frameworks seems not to be useful.

Music in the Indian subcontinent has been a source of aesthetic delight from time immemorial. From the time of Bharata's Natyashastra [10], there have been a number of treatises which speak in favor of the various *rasas* (emotional experiences) that are conveyed by the different forms of musical performances. The aim of any dramatic performance is to emote in the minds of audience a particular kind of aesthetic experience, which is described as "Rasa". The concept of "Rasa" is said to be the most important and significant contribution of the Indian mind to aesthetics [1][2]. Thus, the response of human mind to Hindustani Classical Music demands a different cognitive engagement compared to other forms of music [11].

Music impacts brain in different parts of brain in varied fashion. Music influences and manipulate human emotions and thinking which gives rise to influential music cognition. Studying human brain responses gives excellent insight into understanding of brain. The brain signal analysis gives excellent insight into neural and functional architecture of brain functions. Such studies of brain signal analysis pave structured foundation in the domains of Brain computer interface, music cognition. Such analysis does impact understanding of brain emotions in musical ambience. Advancing computing techniques have proved to be a great thrust in the area of research in music cognition.

A number of issues like universality, arousal-valence and its linkage with modularity or neural correlates have been baffling scientists for quite some time. [3][4][5]. EEG or Electroencephalogram is used for interpretation and analysis of different complex activity of brain signal by attaching efficient bio-sensors to head. It is basically a time versus amplitude plot which contains a lot of information about cerebral cortex nerve of human brain. EEG spectrum can be divided in different frequency band (1) delta ($\delta$) :0-4 Hz (2) theta($\theta$) :4-8 Hz (3) alpha($\alpha$) :8-13 Hz(4) beta($\beta$ ):13-30 Hz (5) gamma($\gamma$): 30-50 Hz. Different types of music has different effects on these bands in different lobes of the brain. Classical music has more pronounce effect on alpha waves, whereas beta waves are more affected by hard rock music [11, 12]. Earlier studies report that alpha power decreases in the left frontal lobe due to pleasant music but it decreases in the right frontal lobe due to unpleasant music [13]. It can be seen that band responses are very much dependent on the type and genre of audio clips being played. Most of the earlier works deal with linear features of EEG signal, which eventually leads to a huge loss of data and involves a lot of approximations. Hence, the use of non-linear analysis to extract features from complex EEG signals is much more advisable.

In this study, we have applied two novel techniques - Detrended Fluctuation Analysis (DFA) and Adaptive Fractal Analysis (AFA) on complex EEG signals obtained under the influence of a pair of Hindustani *raga* clips of contrast emotion. DFA was first used by Peng et.al [14] to determine the long range correlations present in DNA nucleotides. Detrending involves the isolation of the low frequency variation (i.e. trend) and to decompose the residual signal into a seasonal (or cyclic) and a random walk type variation (i.e. white noise or high frequency noise). We use DFA method for the analysis and elimination of trends from data sets. It has successfully been applied to diverse fields such as DNA sequences [14, 15], climate [16], EEG data [17, 18], economical time series [19] etc. Another method to determine the Hurst exponent has been used recently[20], called Adaptive Fractal Analysis (AFA), which utilizes an adaptive detrending algorithm to extract globally smooth trend signals from the data and has shown good applications in various real world data [21, 22]. AFA and DFA techniques have similarity in many aspects, but AFA can deal with arbitrary strong non-linear trends while DFA cannot. We compare the Hurst Exponents obtained from the EEG signals using DFA and AFA techniques under the effect of different musical stimuli. To the best of our knowledge, no similar work has been reported on comparative study of AFA and DFA response for emotion categorization corresponding two contrasting Hindustani classical musical stimuli.

The rest of the paper has been organized as follows: Section II depicts the details about the Theory of AFA and DFA for normal signals. Materials and data used for this work is described in section III. Result table and graphs are shown in section IV.

## II. Theory

### A. Adaptive Fractal Analysis:

AFA or Adaptive Fractal analysis utilizes a detrending algorithm to extract globally smooth trend of signals from data[6]. AFA has the ability of dealing with arbitrary and strong non-linear trends which the DFA method cannot achieve. AFA directly interprets spectral energy in a short time series data on the other hand DFA does not.

The first step in AFA is to identify a globally smooth trend signal that is created by patching together local polynomial fits to the time series. This means a recreation of local features in the data in attempted using simple polynomial functions. The original set of data is divided into windows of length $w = 2n + 1$ with the windows overlapping $n + 1$ points. In each window a best fitting polynomial is fitted of order M which is identified by least square regression method. AFA is practically different from DFA as the AFA is a result of stitching all these local fits to obtain a global smooth trend and not in DFA.

In fact, the scheme ensures that the fitting is continuous everywhere, smooth at the non-boundary points, and has the right and left derivatives at the boundary.

Detrending is done on the data by removing the global trend signal. Detrending method on signal is done by locally varying window. The residual of fit is identified through subtracting global trend from the original series. This is similar to the DFA but in DFA stitching is not performed whereas stitching is must for AFA analysis.

Next step is to examine the relation between variance of magnitude of residuals, F(w) and window size w.

$$F(w) = \left[\frac{1}{N}\sum_{i=1}^{N}\left(u(i) - v(i)\right)^2\right]\hat{}0.5\sim w\hat{}H$$

$w$ is size of the window, u is original function, v is the global trend. $H$ is the Hurst exponent which is signified by the plot of $\log_2 F(w)$ as a function of $\log_2 w$.

### B. Detrended Fluctuation Analysis:

Detrended Fluctuation Analysis has proven to be benificial in understanding unique insights into neural organization[7]. DFA for a time series say {t1, t2, t3,….., tn} can be computed by following:

1. Another series T as [T(1); T(2); T(3)......T(N)] , $T(k)=\sum_{i=1}^{k}(t_i - t_{mean})$. $t_{mean}$ denotes mean of the points in the series t.
2. The series T is under interest. Series T is sliced into threads of length N. Each thread must contain this same number of element which is N. For each of the N element thread, a line is fit which signifies the trend in the thread. The fit is called Tn(k).
3. The detrending is $[T(k) - T_n(k)]^2$ which helps in calculation of RMS fluctuation. $F(N)=\sqrt[2]{\frac{1}{n}\sum_{k=1}^{n}[T(k) - T_n(k)]^2}$ is called root mean square fluctuation.
4. $F(n) \propto n^a$, a is expressed as the slope of logarithimic plot of log[F(n)] versus log(n).

Obtained a is the DFA value of a signal. It is called the DFA scaling exponent and it quantifies self-similarity and correlation properties of time series. As it suggests, a time series having has higher DFA scaling exponent is a symbolic quantification of presence of long range correlation. DFA quantifies complexity of using fractal property.

The scaling exponent a denotes the following:

1. a < 0.5 anti correlated
2. a = 0.5 white noise
3. a > 0.5 positive autocorrelation
4. a = 1 1/f noise
5. a = 1.5 Brownian noise

The DFA exponent can completely signify auto correlation of a signal. DFA technique was applied following the NBT algorithm used in Hardstone et.al [8].

## III. Materials and Methodology

### A. Subjects Summary:

In this study five (5) musically untrained volunteers (M=3, F=2) have participated. The average age was 23 years and average body weight was 70kg. Informed consent was obtained from each subject according to the guidelines of the Ethical Committee of Jadavpur University. The experiment was conducted in the afternoon with a normal diet in a normally conditioned room sitting on a comfortable chair and performed as per the guidelines of the Institutional Ethics Committee of SSN College for Human volunteer research. This is a pilot study, part of an ongoing project in the Centre, the number of subjects will be increased in due course of time; the initial findings have been reported here.

### *B. Experimental Details:*

EEG recording was done with two *ragas – Mia ki Malhar* and *Bahar* of Hindustani music for the 5 subjects. From the complete playing of the ragas, segments of about 2 minutes were cut out for analysis of each raga. The emotional part of each clip was identified with the help of experienced musicians as well as standardized with a listening test of about 50 participants. Listening test data revealed that 73% of the participants found Bahar to be joyful, 85% found Mia ki Malhar to convey sad emotion. These findings were corroborated with the brain response data.

Each of these sound signals was digitized at the sample rate of 44.1 KHZ, 16 bit resolution and in a mono channel. A sound system (Logitech R _ Z-4 speakers) with high S/N ratio was used in the measurement room for giving music input to the subjects.

### *C. Experimental Protocol:*

EEG was done to record the brain-electrical response of 5 subjects. Each subject was prepared with an EEG recording cap with 19 electrodes (Ag/AgCl sintered ring electrodes) placed in the international 10/20 system. Fig. 1 depicts the positions of the electrodes.

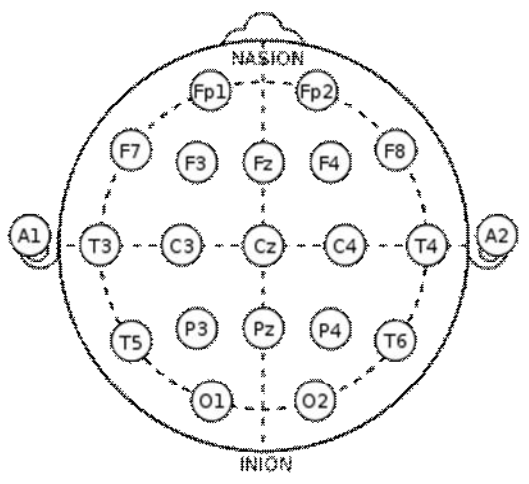

Fig. 1 The position of electrodes according to the 10-20 international system

Impedances were below 5 kOhms. The EEG recording system was operated at 256 samples/s recording on customized software of RMS. The data was band-pass-filtered between 0.5 and 35Hz to remove DC drifts and suppress the 50Hz power line interference. Each subject was seated comfortably in a relaxed condition in a chair in a shielded measurement cabin. They were asked to close eyes. A 12 min recording period was started, and the following protocol was followed:

1. 2 min No Music
2. 2 min With Drone
3. 2 min With Music 1 ( Bahar )
4. 2 min No Music
5. 2 min With Music 2 (Mia ki Malhar)
6. 2 min No Music

The Drone was taken to create baseline over which the effect of other musical clips can be monitored.

### *D. Methodology:*

In order to eliminate all frequencies outside the range of interest, data was band pass filtered with a 0.5-35 Hz FIR filter. DFA, AFA exponent were obtained for 'before music', 'with music' as well as 'without music' conditions for total 6 electrodes F3, F4, O1, O2, T3, T4. The alpha, theta as well as gamma scaling exponent were computed for each of the experimental conditions, but in this study only the effect in the alpha frequency range is reported as it has been reported to be the most relevant when it comes to the facilitating emotional cognition in human brain.

## IV. RESULT

The following (fig. 2-3) are the plots for Adaptive Fractal Analysis (AFA) and Detrended Fluctuation Analysis (DFA) of a particular time series where $R^2$ denotes the sum of the squares of residuals. The smaller values of $R^2$ in both the cases indicate tight fits to the respective data.

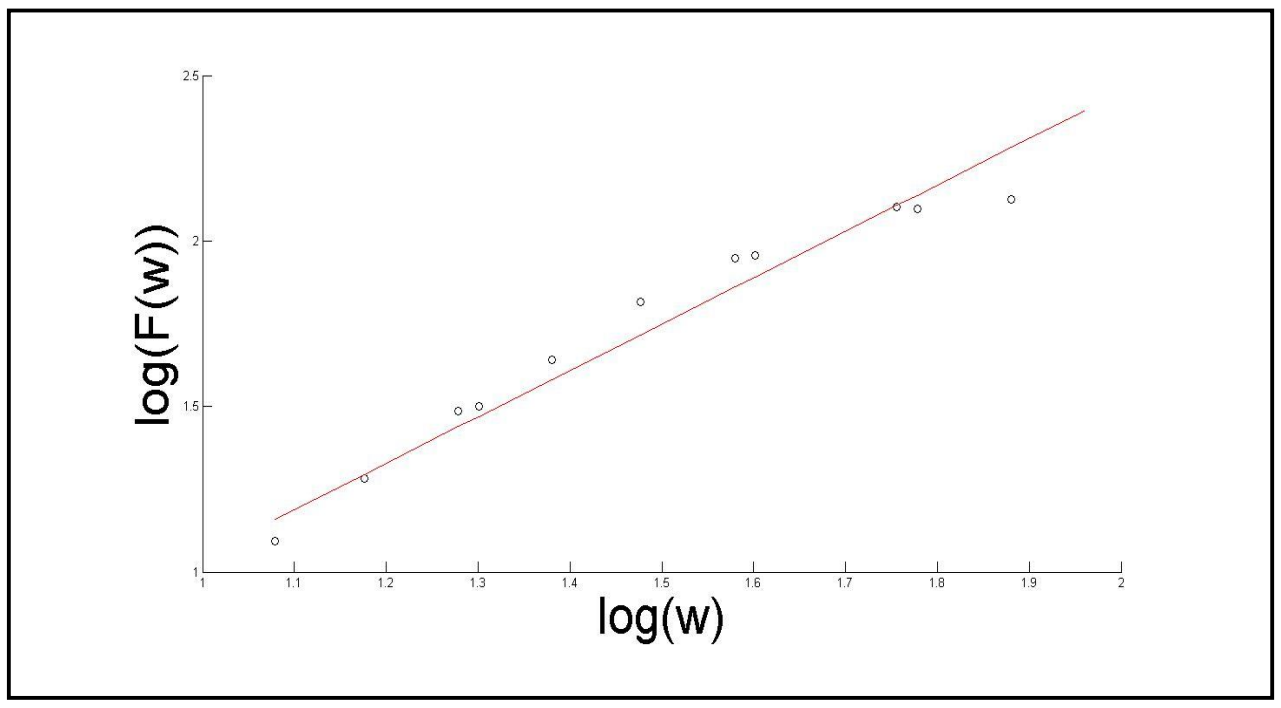

Fig 2: Sample AFA plot with $R^2$=0.0586

Fig 3: Sample DFA plot with $R^2$=0.0049

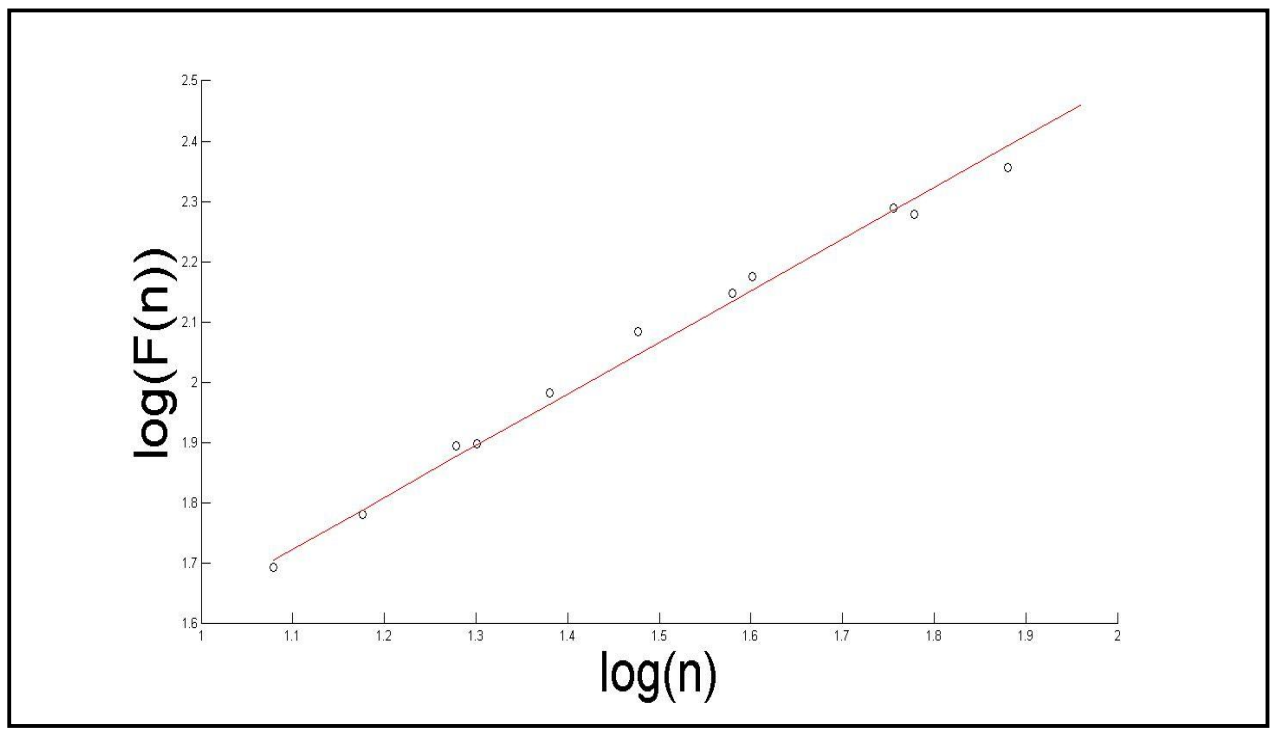

The results for the variation of Hurst Exponent in alpha frequency band under the effect of different musical stimuli is given in the following figures:

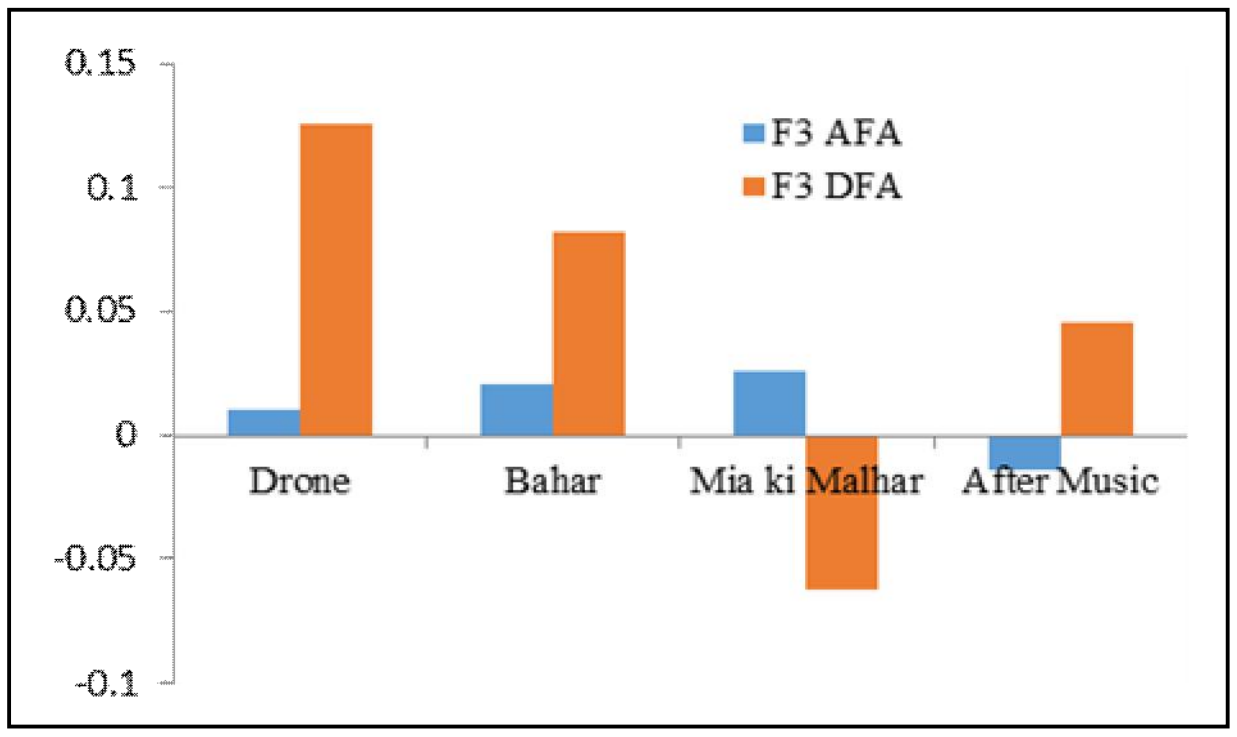

Fig 4: AFA and DFA plot for F3 electrode

In figure 4 the variation of AFA and DFA is plotted for two different ragas and the previous and post condition . Here we can see that there is a major variation in DFA value for the F3 electrode but for AFA the response is almost same. So we can conclude that for DFA there is arousal of hurst exponent for Bahar and suppression for Mia ki Malhar. So two contrasting emotion oriented ragas give a contrasting response for left frontal electrode.

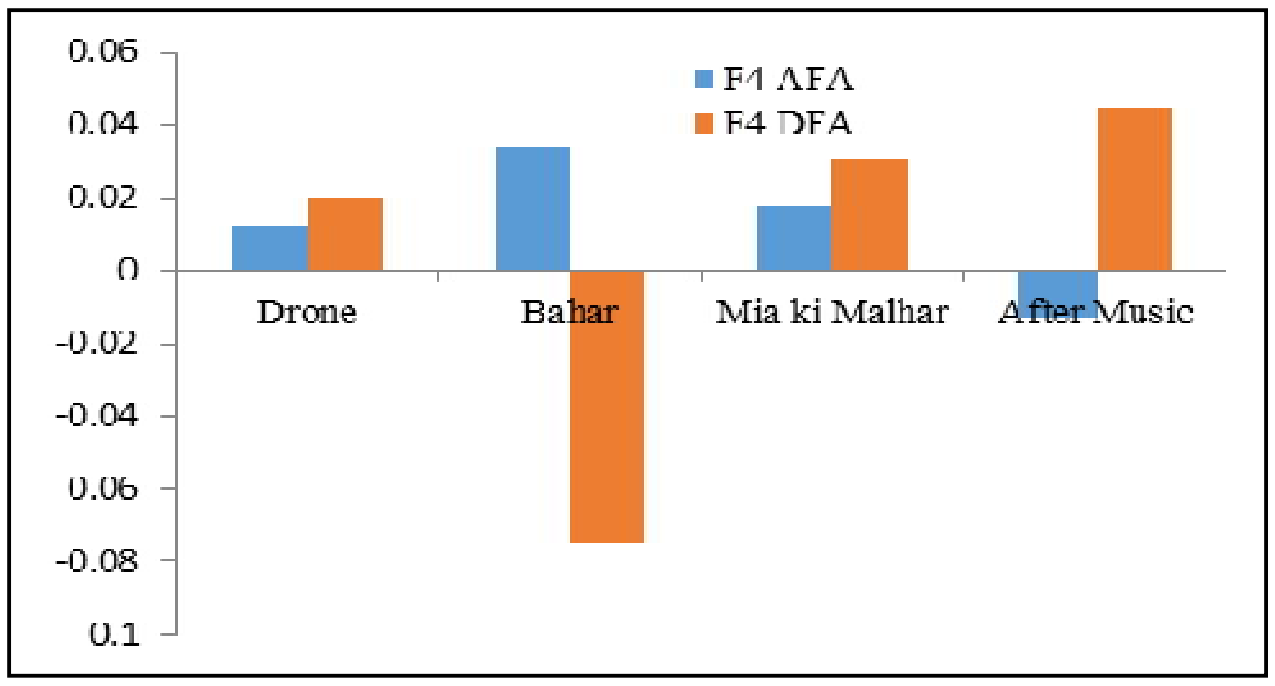

Fig 5: AFA and DFA plot for F4 electrode

For the right frontal electrode F4, however we see that the DFA scaling exponent increases for both the clips, the rise being more significant for the sad clip, clearly indicating the fact that right hemisphere is much more aroused in case of sad clip. Another interesting observation is that after the removal of both the clips there is a significant dip in the values of Hurst exponent.

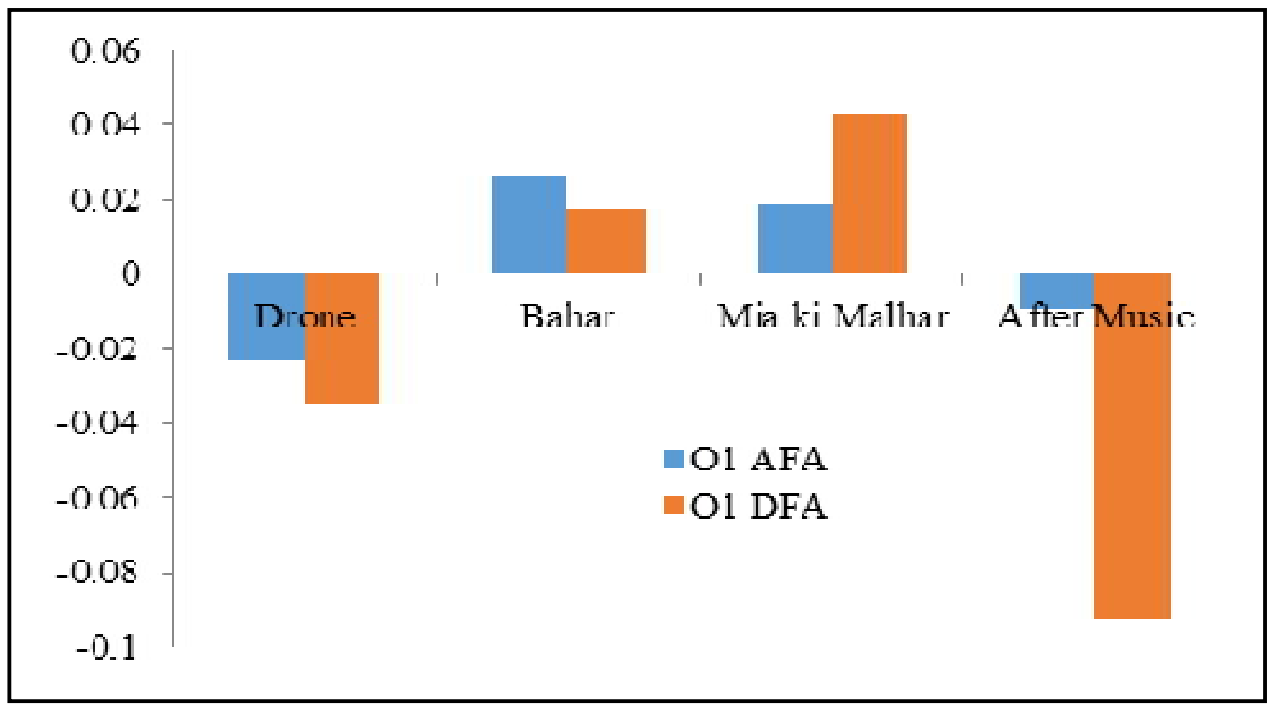

Fig 6: AFA and DFA plot for O1 electrode

This is the plot for left occipital electrode O1 for both DFA and AFA. Here, we see a prominent increase in the Hurst Exponent value under the effect of negative emotional clip while both AFA and DFA scaling exponent falls without music. Though occipital lobe is mainly associated with visual cognition, we see that there is significant response in the auditory domain also.

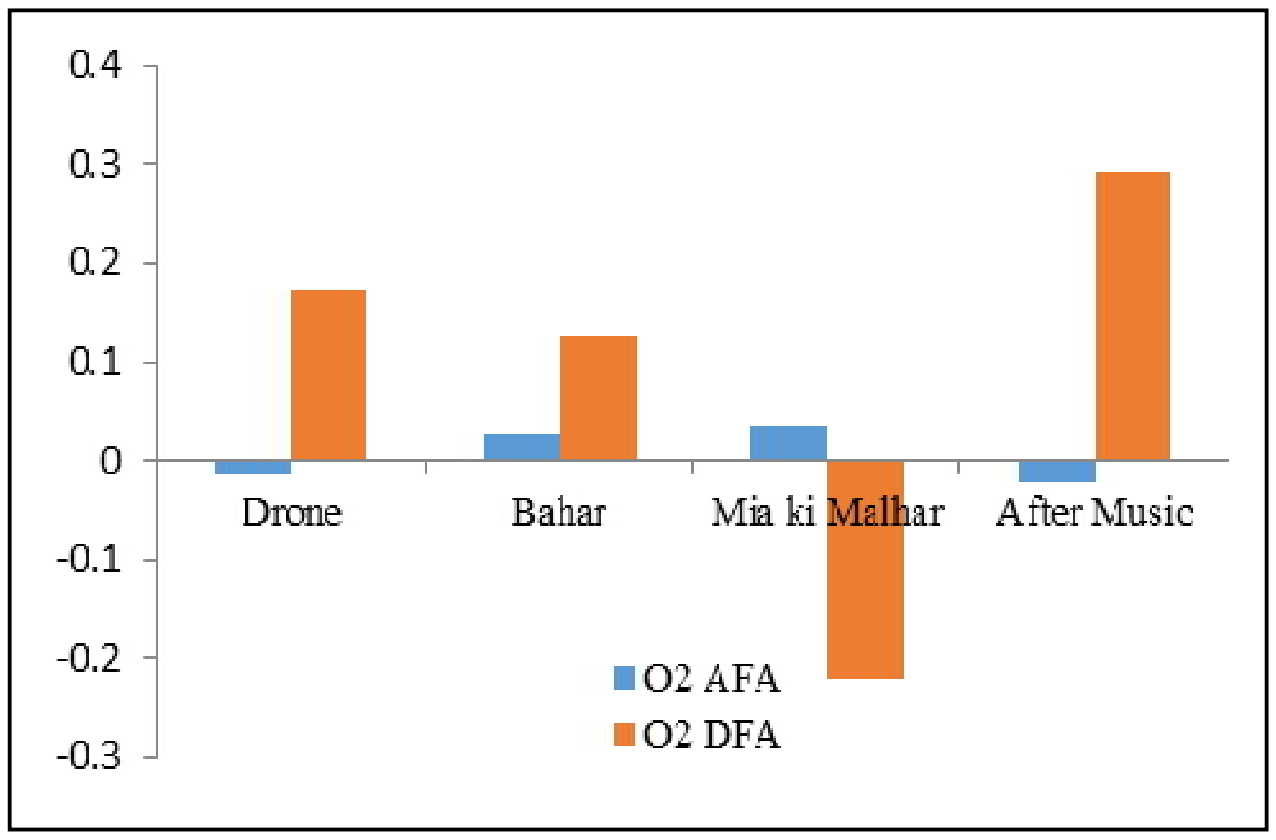

Fig 7: AFA and DFA plot for O2 electrode

For the right occipital lobe O1, response is shown in figure 7, DFA scaling exponent show significant decrease under the effect of sad clip, while there is an increase in case of happy clip. Another interesting observation here is that in the after music condition, significant rise in DFA scaling exponent is noticed. Hurst Exponents determined from AFA technique does not exhibit any considerable change under the effect of emotional stimuli in EEG data.

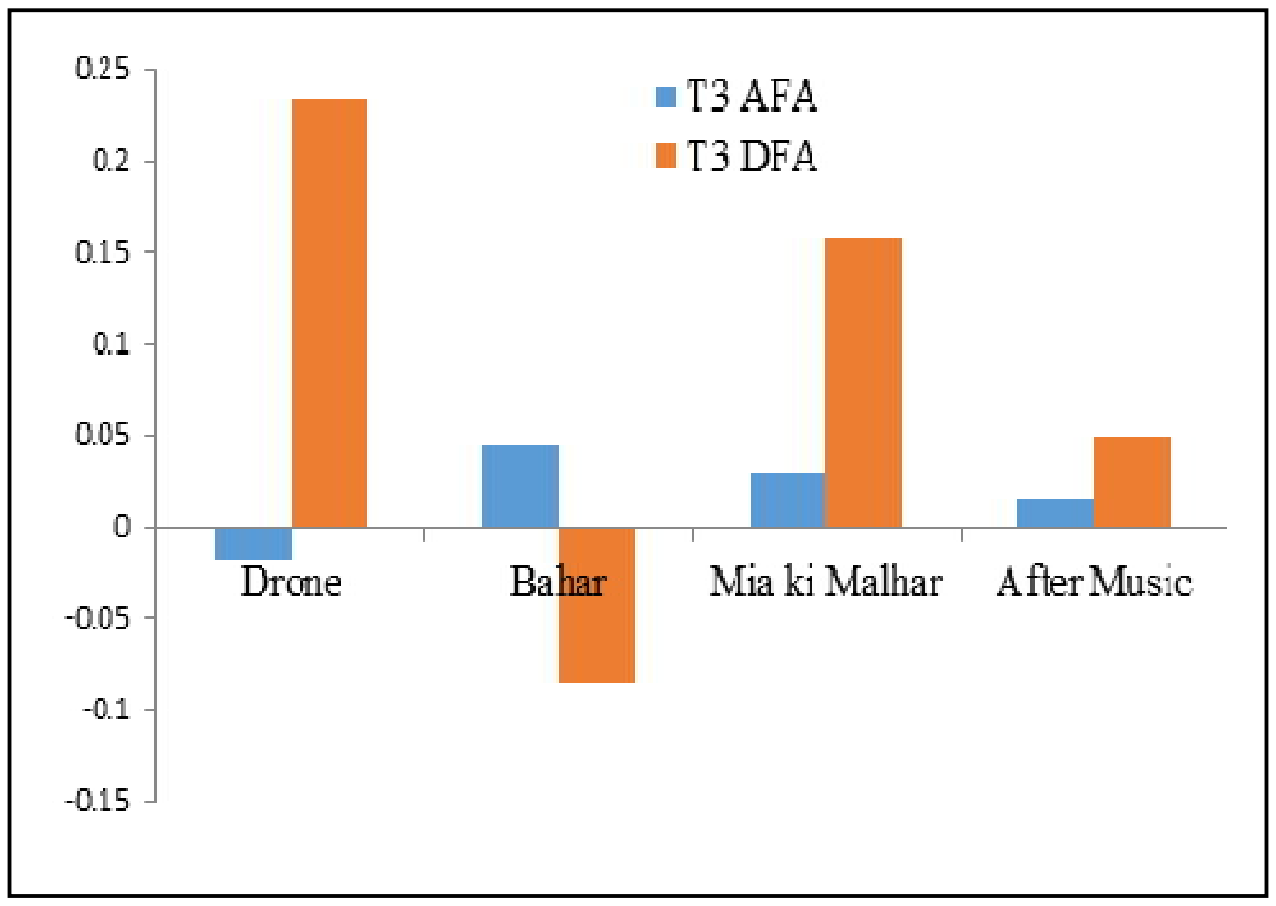

Fig 8: AFA and DFA plot for T3 electrode

In the left temporal electrode (T3) also, response of AFA is quite similar through all the regions. Since temporal lobe is associated with auditory stimuli, we see significant arousal even under the effect of drone. For this electrode response for DFA in happy emotion based clip Bahar suppresses,but sorrow emotion based raga Mia ki Malhar arouses. This observation again points in the direction of categorization of emotional clips using DFA exponent.

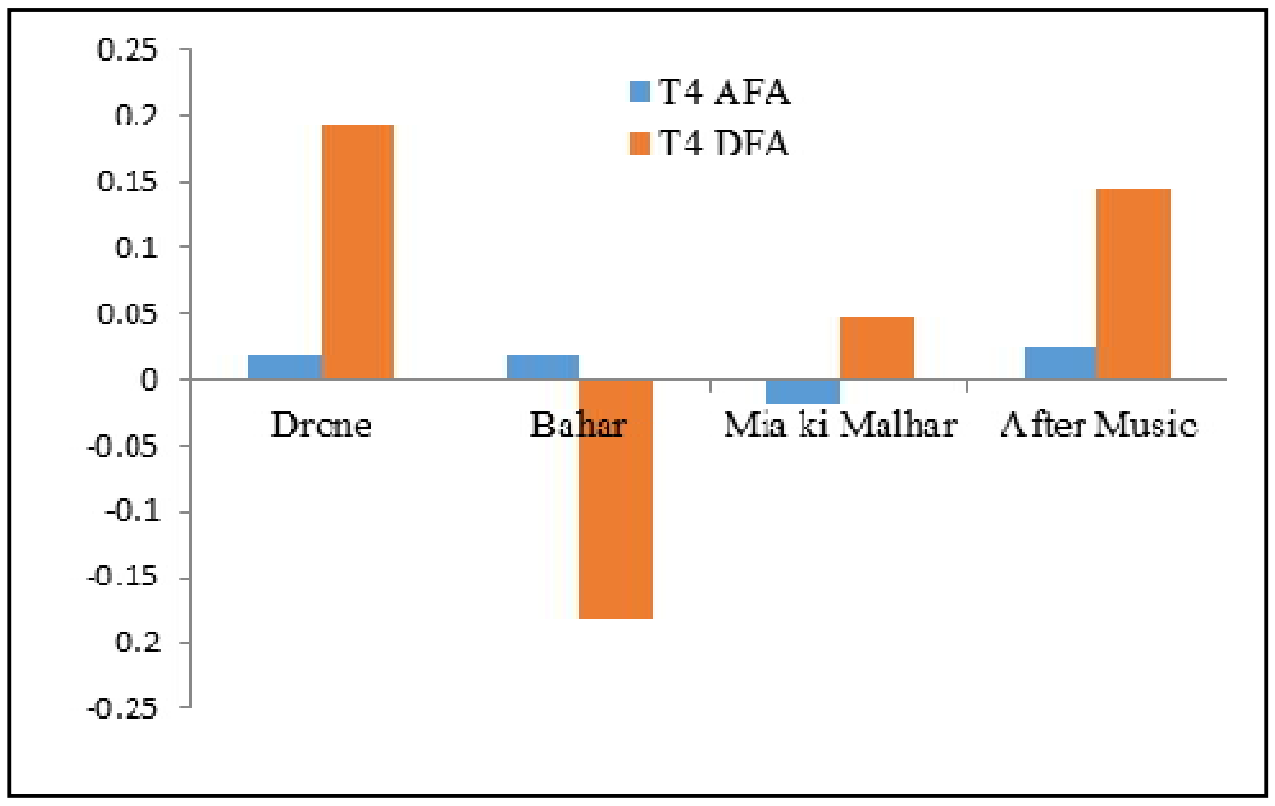

Fig 9: AFA and DFA plot for T4 electrode

For the right Temporal lobe electrode (T4) AFA response is also quite similar in all the cases, though there is a slight change from Bahar to Mia ki Malhar. But for DFA this is quite distinguishable i.e. Hurst exponent for sad clip increases while that for happy clip decreases. Again, a significant response is obtained under the effect of drone stimulus in case of T4 electrode.

## V. Conclusion And Future Work

This is a novel comparative study between the behavior of AFA and DFA regarding music cognition. The results point in the direction of categorization and quantification of musical emotion using non-linear analysis techniques. So far the study result is concerned it is found that DFA is more applicable for different cognition of arousal of different emotions using biomedical time series data. In this study, Hurst Exponent obtained from AFA has not been able to distinguish significantly one emotional state from another which has been done much beautifully by DFA technique. AFA technique proves to be much more efficient in tackling real world data where the time series is short. In the future, we will be expanding our study to check the response of this fractal behavior to all the frequency band of EEG, social data, market trend and image analysis also.

## Acknowledgement

One of the authors, AB acknowledges the Department of Science and Technology (DST), Govt. of India for providing (A.20020/11/97-IFD) the DST Inspire Fellowship to pursue this research work. Another author, SS acknowledges the West Bengal State Council of Science and Technology (WBSCST), Govt. of West Bengal for providing the S.N. Bose Research Fellowship Award to pursue this research (193/WBSCST/F/0520/14).

All the authors acknowledge Department of Science and Technology, Govt. of West Bengal for providing the RMS EEG equipment as part of R&D Project (3/2014).

## References

[1] Juslin, Patrik N., and John A. Sloboda. *Music and emotion: Theory and research. Oxford University Press, 2001.*

[2] Bakan, Michael. *World music: Traditions and transformations*. McGraw-Hill Higher Education, 2007.

[3] Hardstone, Richard, et al. "Detrended fluctuation analysis: a scale-free view on neuronal oscillations." *Scale-free Dynamics and Critical Phenomena in Cortical Activity* (2012):.

[4] Sanyal, Shankha, et al. "Chaotic Brain, Musical Mind-A Non-Linear Neurocognitive Physics Based Study." *Journal of Neurology and Neuroscience* (2016).

[5] Sourya Sengupta, Sayan Biswas, Shankha Sanyal, Archi Banerjee, Ranjan Sen- gupta and Dipak Ghosh"Quantification and Categorization of Emotion us- ing Cross Cultural Music: An EEG Based Fractal Study IEEE 2nd International Conference on Next Generation Computing Technology,(NGCT) Dehradun 2016.

[6] Riley, Michael A., et al. "A tutorial introduction to adaptive fractal analysis."*Frontiers in physiology* 3 (2012): 371.

[7] Golińska, Agnieszka Kitlas. "Detrended fluctuation analysis (DFA) in biomedical signal processing: selected examples." Stud. Logic Grammar Rhetoric 29 (2012): 107-115 -

[8] Hardstone, Richard, et al. "Detrended fluctuation analysis: a scale-free view on neuronal oscillations." *Scale-free Dynamics and Critical Phenomena in Cortical Activity* (2012):.

[9] Di Lorenzo P (2002), Chaos structures in Gregorian Chant, Proc. Musical Creativity- 10th Anniversary ESCOM, Liege, Belgium

[10] Martinez, José Luiz. *Semiosis in Hindustani music*. Vol. 15. Motilal Banarsidass Publ., 2001.

[11] Banerjee, A., Sanyal, S., Patranabis, A., Banerjee, K., Guhathakurta, T., Sengupta, R., ... & Ghose, P. (2016). Study on Brain Dynamics by Non Linear Analysis of Music Induced EEG Signals. *Physica A: Statistical Mechanics and its Applications*, *444*, 110-120.

[12] Natarajan, Kannathal, et al. "Nonlinear analysis of EEG signals at different mental states." *BioMedical Engineering OnLine* 3.1 (2004): 1.

[13] Schmidt, L. A., & Trainor, L. J. (2001). Frontal brain electrical activity (EEG) distinguishes valence and intensity of musical emotions. *Cognition & Emotion*, *15*(4), 487-500.

[14] Peng CK, Buldyrev SV, Havlin S, Simons M, Stanley HE, Goldberger AL. Mosaic organization of DNA nucleotides. Physical review e. 1994 Feb 1;49(2):1685.

[15] Buldyrev, S. V., et al. "Analysis of DNA sequences using methods of statistical physics." *Physica A: Statistical Mechanics and its Applications* 249.1 (1998): 430-438.

[16] Ivanova K, Ausloos M. Application of the detrended fluctuation analysis (DFA) method for describing cloud breaking. Physica A: Statistical Mechanics and its Applications. 1999 Dec 1;274(1):349-54.

[17] Banerjee A, Sanyal S, Patranabis A, Banerjee K, Guhathakurta T, Sengupta R, Ghosh D, Ghose P. Study on Brain Dynamics by Non Linear Analysis of Music Induced EEG Signals. Physica A: Statistical Mechanics and its Applications. 2016 Feb 15;444:110-20.

[18] Sanyal S, Banerjee A, Pratihar R, Maity AK, Dey S, Agrawal V, Sengupta R, Ghosh D. Detrended Fluctuation and Power Spectral Analysis of alpha and delta EEG brain rhythms to study music elicited emotion. InSignal Processing, Computing and Control (ISPCC), 2015 International Conference on 2015 Sep 24 (pp. 205-210). IEEE.

[19] Wang Y, Liu L. Is WTI crude oil market becoming weakly efficient over time?: New evidence from multiscale analysis based on detrended fluctuation analysis. Energy Economics. 2010 Sep 30;32(5):987-92.

[20] Riley, Michael A., et al. "A tutorial introduction to adaptive fractal analysis." *Frontiers in physiology* 3 (2012).

[21] Gao, J., Hu, J., Mao, X., & Perc, M. (2012). Culturomics meets random fractal theory: insights into long-range correlations of social and natural phenomena over the past two centuries. *Journal of The Royal Society Interface*, *9*(73), 1956-1964

[22] Kuznetsov, Nikita, et al. "Adaptive fractal analysis reveals limits to fractal scaling in center of pressure trajectories." *Annals of biomedical engineering*41.8 (2013): 1646-1660.